\documentclass[prd,showpacs,preprintnumbers,nofootinbib]{revtex4}

\usepackage{amsmath,amssymb}
\usepackage{bm}
\usepackage{mathrsfs}
\usepackage{braket}
\usepackage[dvipdfmx]{graphicx}

\begin{document}
\preprint{CHIBA-EP-217-v2, 2016}

\title{
Gauge-covariant decomposition and magnetic monopole for G(2) Yang-Mills field
}

\author{Ryutaro Matsudo}
\email{afca3071@chiba-u.jp}

\author{Kei-Ichi Kondo}
\email{kondok@faculty.chiba-u.jp}

\affiliation{Department of Physics,  
Graduate School of Science, 
Chiba University, Chiba 263-8522, Japan
}
\begin{abstract}
We give a gauge-covariant decomposition of the Yang-Mills field with an exceptional gauge group $G(2)$, which extends  the field decomposition invented by Cho, Duan-Ge, and Faddeev-Niemi  for the $SU(N)$ Yang-Mills field. 
As an application of the decomposition, we derive a new expression of the non-Abelian Stokes theorem for the  Wilson loop operator in an arbitrary representation of $G(2)$. 
The resulting new form is used to define gauge-invariant magnetic monopoles in the $G(2)$ Yang-Mills theory. Moreover, we obtain the quantization condition to be satisfied by the resulting magnetic charge.
The method given in this paper is general enough to be applicable to any semi-simple Lie group other than $SU(N)$ and $G(2)$. 
\end{abstract}


\maketitle

\section{Introduction}

Understanding the mechanism underlying quark confinement from the first principle of QCD  is still a challenging problem in theoretical particle physics \cite{Wilson74}. 
As a possible step towards this goal, it will be efficient to extract the dominant field mode $\mathscr{V}$ responsible for confinement from the Yang-Mills field $\mathscr{A}$ to clarify the physics behind the phenomena of confinement. 
The well-known mathematical identity called the Cartan decomposition \cite{Gilmore06} is used to decompose the  field variable $\mathscr{A}$ valued in the Lie algebra  $\mathscr{G}={\rm Lie}(G)$ of a gauge group $G$ into the  simultaneously diagonalizable part in the Cartan subalgebra $\mathscr{H}={\rm Lie}(H)$ and the remaining off-diagonal part in
the orthogonal complement of ${\rm Lie}(H)$.
However, the Cartan decomposition is not suited for studying the non-perturbative features of the gauge field theory with local gauge invariance, since the Cartan decomposition cannot retain the original form after the gauge transformation, namely, the local rotation of the Cartan-Weyl basis for the Lie algebra. 

In view of these, the novel decomposition called the  Cho-Duan-Ge-Faddeev-Niemi (CDGFN) decomposition \cite{DG79,Cho80,FN98,Shabanov99,KMS06,KMS05,Kondo06,Cho80c,FN99a,BCK02,KSM08,KKSS15} is quite attractive, since the CDGFN decomposition given in the form $\mathscr{A}=\mathscr{V}+\mathscr{X}$ is gauge covariant, namely, it keeps its form under the gauge transformation or the local color rotation.
In the CDGFN decomposition, the unit Lie algebra valued field $\bm{n}_{j}$ called the color direction field or the color field for short plays the crucial role for retaining the local gauge covariance of the decomposition. 
For $G=SU(N)$, the color field $\bm{n}_{j}(x)$ is constructed from the maximally commuting generators $H_j$ $(j=1,..., {\rm rank}G)$ in the Cartan subalgebra $\mathscr{H}={\rm Lie}(H)$ according to the local adjoint rotation by a group element $g $ of the gauge group $G$ at every point $x$ of spacetime:
\begin{align}
 \bm{n}_{j}(x) = g(x) H_j g^\dag(x) 
 , \ g(x) \in G .
\end{align}
The color direction field belongs to the subset of ${\rm Lie}(G)$ which is topologically equivalent to $G/\tilde{H}$,
where a subgroup $\tilde{H}$ called the maximal stability subgroup of $G$ is specified from the degeneracy among the eigenvalues of a representation matrix of $H_j$. 
In other words, the color field $\bm{n}_{j}$ is regarded as the local embedding of the Cartan direction $H_j$ in the internal space of the non-Abelian group $G$.  
From this viewpoint, the Cartan decomposition is identified with a global limit of the CDGFN decomposition, which urges us to consider that the Abelian projection method \cite{tHooft81} is nothing but a gauge-fixed version of the gauge-covariant CDGFN decomposition. 
The application of the novel decomposition to the Yang-Mills non-Abelian gauge field paves the way for understanding quark confinement in a gauge-independent manner. 
In fact, this method has been extensively used to investigate quark confinement in the $SU(N)$ Yang-Mills theory in the last decade, see e.g., \cite{KKSS15} for a review.


A promising mechanism for understanding quark confinement is well known as the dual superconductivity \cite{dualsuper}. 
It is a hypothesis based on the electro-magnetic dual analog of the type II superconductor in which the magnetic field applied to the bulk of the superconductor is squeezed to form the magnetic vortex due to the Meissner effect of excluding the magnetic field  from the superconductor \cite{superconcuctor}. 
The color electric field created by a pair of a quark and an antiquark would be squeezed to form an electric flux-tube or a hadron string with its ends on a quark and an antiquark.  
For the dual superconductivity to work, therefore, one needs magnetic objects, say magnetic monopoles to be condensed in the vacuum of the Yang-Mills theory, which is supposed to be dual to the ordinary superconductivity caused by condensation of electric objects, a pair of electrons called the Cooper pairs \cite{superconcuctor}. 

The magnetic monopole in the pure Yang-Mills theory has been mostly constructed by the Abelian projection method, which  \cite{tHooft81} breaks explicitly the original non-Abelian gauge group $G$ to the maximal torus subgroup $H$. 
However, this is not the only way to define magnetic monopoles in pure Yang-Mills theory without the Higgs field. 
In fact, one can define gauge-invariant magnetic monopoles in the pure Yang-Mills theory without breaking the original gauge symmetry by using the non-Abelian Stokes theorem  \cite{DP89,DP96,KondoIV,KT00b,KT00,Kondo08,Kondo08b,MK15} for the Wilson loop operator which is in itself gauge invariant \cite{Wilson74}. 
The gauge-invariant magnetic monopole is specified by the maximal stability subgroup $\tilde H$ which is uniquely determined for the highest-weight state of a given representation for a quark source. 
For quarks in the fundamental representation of $SU(N)$, especially, the maximal stability subgroup $\tilde H$ is given by $\tilde H =U(N-1)$, which is distinct from the maximal torus group $U(1)^{N-1}$ for $N\geq3$. 

The contribution of magnetic monopoles to the Wilson loop average can be calculated by using the path-integral framework using the reformulation of the Yang-Mills theory based on change of variables in accord with the field decomposition. 
We find that there are some options for reformulating the $SU(N)$ Yang-Mills theory ($N\geq3$) corresponding to the different choice of the color direction field compatible with the maximal stability group, although there is only one way to reformulate the $SU(2)$ Yang-Mills theory.
Whichever options we use, we need the field decomposition formula, which allows us to decompose an arbitrary element $\mathscr{F}$ of a Lie algebra $\mathscr{G}$ to the part $\mathscr{F}_{\tilde{H}}$ in the Lie algebra of   $\tilde{H}$ and the remaining part $\mathscr{F}_{G/\tilde{H}}$.
See e.g., \cite{KSM08,Kondo08} and also \cite{KKSS15} for a review.

Another popular object of topological nature which is believed to be responsible for confinement is the center vortex \cite{center-vortex,Cornwall}, which is associated to the center subgroup of $G$.  
In fact, the confinement/deconfinement phase transition at finite temperature in the $SU(N)$ Yang-Mills theory is associated with the restoration/spontaneous breaking of the center symmetry $Z(N)$, which is signaled by vanishing/nonvanishing of the Polyakov loop average. 
See e.g., \cite{Greensite} for reviews. 
We suppose, however, that magnetic monopoles and magnetic vortices cannot be independent topological objects. 
They could be different views of a single physical object, just like two sides of a coin, to be simultaneously defined in a self-consistent way \cite{Kondo08b,KKSS15}.
However, this statement remains still a conjecture to be proved. 

The purpose of this paper is to extend the   gauge-covariant field decomposition of the Yang-Mills field  and the non-Abelian Stokes theorem for the Wilson loop operator developed so far for  $SU(N)$ to the exceptional group $G(2)$, which is a preliminary step toward reformulating the $G(2)$ Yang-Mills theory \cite{HMPW03} to discuss the mechanism for confinement/deconfinement. 
Our interests of the exceptional group $G(2)$ lie in a fact that the $G(2)$ Yang-Mills theory has the linear potential \cite{GLORT07,WWW11,BCPP15} and that the center vortex confinement mechanism is argued to work for $G(2)$ in \cite{GLORT07}, although
  $G(2)$ has a trivial center subgroup, consisting only of the identity element \cite{GG73,CCDO05}. 
We want to define the gauge-invariant  magnetic monopole in the $G(2)$ Yang-Mills theory and then  examine whether the magnetic monopole defined in our framework can be a universal topological object responsible for confinement, irrespective of the gauge group. 
This investigation will help us to prove or disprove the above conjecture on the interrelation between magnetic monopoles and magnetic vortices. 

The present paper is organized as follows. 
In sec. II, we first  determine the maximal stability subgroup for a given representation of $G(2)$, 
after presenting some basic properties of $G(2)$. 
In sec. III, we subsequently derive the gauge-covariant decomposition formula for the $G(2)$ Yang-Mills field corresponding to each maximal stability subgroup. 
We show that the $G(2)$ Yang-Mills field has different  gauge-covariant decompositions depending on the maximal stability subgroup, which are more complicated than those obtained for the $SU(N)$ Yang-Mills field. 
In fact, it turns out that the decomposition formula for $G(2)$ cannot be obtained as a simple extension of that for $SU(N)$.
This is because  the fact that all roots of $SU(N)$ have the same norm was used in deriving the decomposition formula for $SU(N)$. 
However, some roots of $G(2)$ have different norm from  the other roots.
Consequently, the relevant decomposition for  $G(2)$ cannot be obtained by using the double commutators, in sharp contrast to $SU(3)$.
Nevertheless, the multiple commutators with the Cartan generators enable us to obtain the desired decomposition.
Remarkably, we have found that the decomposition formula for $G(2)$ can be obtained  using sextuple commutators with the Cartan generators or the color fields.
Moreover, the method presented in this paper for obtaining the decomposition formula for $G(2)$ can be applied to any semi-simple Lie algebra, and therefore the reformulation of the Yang-Mills theory would be possible for an arbitrary semi-simple gauge group.

In sec. IV, we derive a non-Abelian Stokes theorem for the Wilson loop operator of the $G(2)$ Yang-Mills field, which is written using the sextuple commutators with the color direction fields, as an application of  the decomposition formula. 
This enables us to define gauge-invariant magnetic monopoles in the $G(2)$ Yang-Mills theory in sec. V. 
We find that which kind of magnetic monopoles can be defined is determined by the stability subgroup of $G$. 
We show that the magnetic charge derived from the gauge-invariant magnetic monopole is subject to a novel quantization condition, which is  similar to, but different from the  quantization condition for the Dirac magnetic monopole and 'tHooft-Polyakov magnetic monopole.
The final section is devoted to conclusion and discussion. 
Some technical derivations are collected in  Appendices A, B and C.

\section{An exceptional group $G(2)$}

\subsection{Basic properties of $G(2)$}

\begin{figure}[t]
\begin{center}
\includegraphics[width=0.1\hsize]{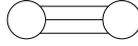}
\caption{The Dynkin diagram of $G(2)$.}
\label{fig:dynkin}
\end{center}
\end{figure}

\begin{figure}[t]
\begin{center}
\includegraphics[width=0.3\hsize]{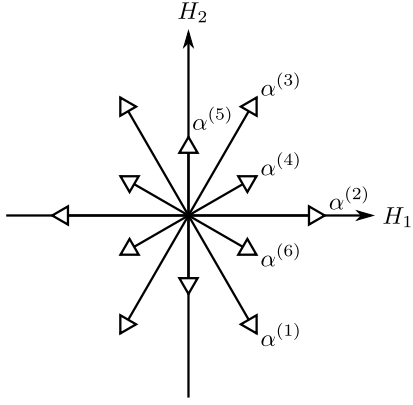}
\caption{The root diagram of $G(2)$.}
\label{fig:root}
\end{center}
\end{figure}

In this section, we give some basic properties of an exceptional group G(2).
We begin with the Dynkin diagram of $G(2)$ given by FIG.~\ref{fig:dynkin}.
It indicates that $G(2)$ has two simple roots (i.e., rank 2) with the opening angle $5\pi/3$.
In this paper, we use
\begin{align}
	\alpha^1 &= (\frac12,-\frac{\sqrt3}2) =: \alpha^{(1)}, \notag\\
	\alpha^2 &= (0,\frac1{\sqrt3}) =: \alpha^{(5)},
\end{align}
as simple roots. 
We see that the other positive roots are obtained as
\begin{align}
	\alpha^1+\alpha^2 &= (\frac12,-\frac1{2\sqrt3}) =: \alpha^{(6)}, \notag\\
	\alpha^1+2\alpha^2 &= (\frac12,\frac1{2\sqrt3}) =: \alpha^{(4)},\notag\\
	\alpha^1+3\alpha^2 &= (\frac12,\frac{\sqrt3}2) =: \alpha^{(3)},\notag\\
	2\alpha^1+3\alpha^2 &= (1,0)=: \alpha^{(2)}.
\end{align}
FIG.~\ref{fig:root} is the root diagram of $G(2)$.
Hence, there are two Cartan generators $H_k$ ($k=1,2$) and twelve shift operators $E_{\alpha}$ ($\alpha\in\mathcal{R}$), where $\mathcal{R}$ is the root system, i.e., the set of positive and negative root vectors.
They satisfy the commutation relation called the Cartan standard form:
\begin{align}
	[H_j, H_k] &= 0 \quad (j,k=1,2), \notag\\
	[H_k, E_\alpha] &= \alpha_k E_\alpha, \notag\\
	[E_\alpha, E_{-\alpha}] &= \alpha\cdot H \notag\\
	[E_\alpha, E_\beta] &\propto \begin{cases}
		E_{\alpha+\beta} & (\alpha+\beta\in \mathcal{R})\\
		0 & ({\rm otherwise})
		\end{cases}
	\label{3}
\end{align}
where $\alpha_k$ denotes the $k$th component of the root vector $\alpha$ and $\alpha\cdot H$ is the inner product defined by $\alpha\cdot H := \alpha^kH_k$.
In this paper we consider a unitary representation. Therefore representation matrices satisfy the Hermiticity:
\begin{align}
	R(H_k)^\dag &= R(H_k), \quad
	R(E_\alpha)^\dag = R(E_{-\alpha}), 
\end{align}
and the normalization:
\begin{align}
	\kappa{\rm tr}(R(H_j)R(H_k)) &= \delta_{jk},\quad
	\kappa{\rm tr}(R(E_{\alpha})R(E_{\beta})) = \begin{cases}
		1 & (\beta = -\alpha) \\
		0 & ({\rm otherwise})
	\end{cases},
\end{align}
where the value of $\kappa$ depends on the representation.
Using this property, we can define the inner product in the Lie algebra as
\begin{align}
	(\mathscr F_1, \mathscr F_2) = \kappa{\rm tr}(R(\mathscr{F}_1) R(\mathscr{F}_2)),
\end{align}
which is independent of the representation $R$.

\begin{figure}[t]
\begin{center}
\includegraphics[width=0.5\hsize]{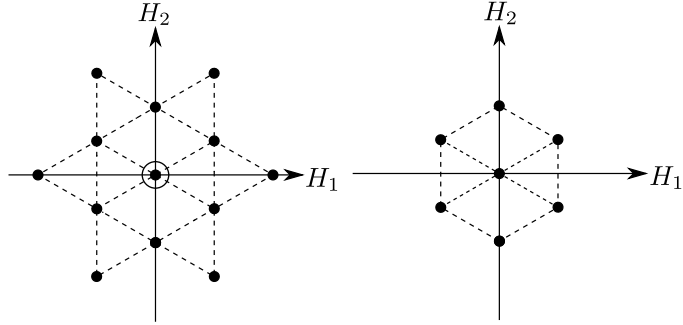}
\caption{The weight diagrams of the fundamental representations.}
\label{fig:diag}
\end{center}
\end{figure}

The weight vector $\vec{\mu}$ of a representation specified by the Dynkin index $[m_1, ...,m_r]$ of the Lie group with the rank $r$ is obtained from the relation:
\begin{align}
 \frac{2\vec{\alpha}^{k} \cdot \vec{\mu}}{\vec{\alpha}^{k} \cdot \vec{\alpha}^{k}} =  m_{k} . 
\end{align}
For $G(2)$, the highest-weight vector $\mu^j$ ($j=1,2$) of a representation with the Dynkin index $[1,0]$ or $[0,1]$ satisfies
\begin{align}
	\frac{2\alpha^k \cdot \mu^j}{\alpha^{k} \cdot \alpha^{k}} = \delta_{kj},
\end{align}
and hence determined as
\begin{align}
	\mu^1 = (1,0), \quad
	\mu^2 = \left( \frac12,\frac1{2\sqrt3} \right) .
\end{align}
The weight diagrams are determined by $\mu^j$ ($j=1,2$), as given in FIG.~\ref{fig:diag}.
The highest-weight vector ${\mu}^{1}$ corresponds to the 14-dimensional adjoint representation $\bm{14}$ with the Dynkin index $[1,0]$, while 
the highest-weight vector ${\mu}^{2}$ corresponds to the 7-dimensional fundamental representation $\bm{7}$ with the Dynkin index $[0,1]$. 
An arbitrary irreducible representation of $G(2)$ is labeled by the two Dynkin indices $[n,m]$ and its highest weight $\Lambda$ can be written as
\begin{align}
	\Lambda &= n\mu^1 + m\mu^2
	= \left( \frac{2n+m}2,\frac m{2\sqrt3} \right) .
\end{align}

Notice that $G(2)$ contains the Lie group $SU(3)$ as a subgroup. We can see from the root diagram that the Lie algebra $su(3)$ of $SU(3)$, denoted as $su(3)={\rm Lie}(SU(3))$, is generated by a set of elements in $su(3)$:
\begin{align}
\label{8}
	\{H_1,H_2,E_{\alpha^{(1)}}, E_{\alpha^{(2)}},E_{\alpha^{(3)}},E_{-\alpha^{(1)}},E_{-\alpha^{(2)}},E_{-\alpha^{(3)}}\} \subset su(3)={\rm Lie}(SU(3)).
\end{align}
Therefore, a representations of $G(2)$ is written as direct sums of representations of $SU(3)$.
For example, the fundamental representations of $G(2)$ are written as
\begin{align}
	\bm{7} &= \bm{3} + \bm{3}^* + \bm{1} ,\\
	\bm{14} &= \bm{8} + \bm{3} + \bm{3}^*.
\end{align}


\subsection{Maximal stability subgroups}

It is known \cite{KSM08,KKSS15} that one can construct a number of the reformulations of the Yang-Mills theory which are discriminated by the maximal stability subgroup.
Therefore it is important to know which subgroup is identified with the maximal stability subgroup for each representation.
In view of this, we first derive a certain property to be satisfied by the generators belonging to the Lie algebra of the maximal stability subgroup of $G(2)$. 
By using this property, then, we determine the maximal stability subgroup for each representation of $G(2)$.

The maximal stability subgroup $\tilde{H}$ for the representation $R$ of a group $G$ is defined to be a subgroup whose element $h\in\tilde{H}$ leaves the highest-weight state $\ket{\Lambda}$ of the representation $R$ invariant up to a phase factor\footnote{
Strictly speaking, we should write
\begin{align}
	R(h)\ket{\Lambda} = \ket{\Lambda}e^{i\phi(h)}, \notag
\end{align}
but if we do so the presentation become rather cumbersome.
Therefore we omit $R(\cdot)$ throughout this subsection.
}:
\begin{align}
	h\ket{\Lambda} = \ket{\Lambda}e^{i\phi(h)} .
\end{align}
Hence, an element of its Lie algebra $\tilde{h}={\rm Lie}(\tilde{H})$
can be written as a linear combination of the Cartan generators and shift-up and -down operators $E_{\alpha}$ and $E_{-\alpha}$ (where $\alpha$ is a positive root) such that $E_{\alpha}\ket{\Lambda} =0$ and $E_{-\alpha}\ket{\Lambda} = 0$.
Here notice that, if there is $E_\alpha$ in the linear combination, then there is also $E_{-\alpha}$, that is to say,  the mutually Hermitian-conjugate generators $E_\alpha$ and $E_{-\alpha}$ must appear in pairs in the linear combination, since all matrices in a unitary representation of the Lie algebra are Hermitian.

We show in the following that $E_{\alpha}\ket{\Lambda}=0$ and $E_{-\alpha}\ket{\Lambda}=0$ if and only if $\Lambda\cdot \alpha =0$.
Here, we should remember that $(\alpha\cdot H)/\alpha^2$ and $E_{\pm\alpha}/|\alpha|$ satisfy the commutation relations of $su(2)$.
We see from this fact that if $\alpha\cdot H\ket{\mu}=0$ and $E_\alpha\ket{\mu}=0$ then $\ket{\mu}$ belongs to the space of the trivial representation of $SU(2)$ and hence $E_{-\alpha}\ket{\mu}=0$.
Because $\ket{\Lambda}$ is highest weight state, $E_\alpha\ket{\Lambda} = 0$.
Hence if $\Lambda\cdot\alpha =0$ then $E_{-\alpha}\ket{\Lambda}=0$.
In the same way, the converse can be proven.

Thus we arrive at the conclusion that $X\in\tilde{h}$ can be written as a linear combination of the Cartan generators $H_j$ and shift operators $E_{\pm\alpha}$ with positive root vectors $\alpha$ that are orthogonal to the highest-weight vector $\Lambda$:
\begin{align}
	\alpha \cdot \Lambda = 0.
\end{align}

Thus, it is easy to see that all representations of $G(2)$ are classified into the following three categories.
\begin{enumerate}
\item
For the highest weight $\Lambda = m\mu^2$, the positive root orthogonal to the highest weight is $\alpha^1=\alpha^{(1)}$ alone.
Hence, the maximal stability subgroup is a $U(2)$ with the generators $H_1$, $H_2$, $E_{\alpha^{(1)}}$ and $E_{-\alpha^{(1)}}$:
\begin{align}
\label{12}
	\tilde{H} = U(2);\; {\rm Lie}(U(2)) = u(2) \supset \{H_1,H_2,E_{\alpha^{(1)}},E_{-\alpha^{(1)}}\} ,
\end{align}
which agrees with a subset of $SU(3)$ specified by (\ref{8}).

\item 
For the highest weight $\Lambda = n\mu^1$, the  positive root orthogonal to the highest weight is $\alpha^2=\alpha^{(5)}$ alone.
Hence the maximal stability subgroup is another $U(2)$ with the generators $H_1$, $H_2$, $E_{\alpha^{(5)}}$ and $E_{-\alpha^{(5)}}$:
\begin{align}
\label{13}
	\tilde{H} = U'(2); \; {\rm Lie}(U'(2)) = u'(2) \supset \{H_1,H_2,E_{\alpha^{(5)}},E_{-\alpha^{(5)}}\} ,
\end{align}
which differs from a subset of $SU(3)$ specified by (\ref{8}).

\item
For the highest weight $\Lambda = n\mu^1 + m\mu^2$ ($n\neq0\neq m$), the maximal stability subgroup is equal to the maximal torus subgroup $U(1)\times U(1)$ generated by the Cartan subalgebra $\{ H_1,H_2\}$:
\begin{align}
\label{14}
	\tilde{H}=U(1)\times U(1);\; {\rm Lie}(U(1)\times U(1)) \supset \{ H_1,H_2\}.
\end{align}
This fact is confirmed as follows.
We can write any positive root as $k\alpha^1 + l\alpha^2$, where $k$ and $l$ are non-negative integers that are not zero simultaneously.
Hence, the relation, 
$
	\Lambda\cdot\alpha  = (n\mu^1+m\mu^2)\cdot(k\alpha^1+l\alpha^2)
	 = nk+ml ,
$
 implies that all positive roots are not orthogonal to the highest weight when $n\neq0$ and $m\neq0$.
Thus, in this case, the generators of the maximal stability subgroup are given by $H_1$ and $H_2$.
\end{enumerate}

\section{Decomposition formula}
\label{sec:decom}

Let $\mathscr{F}$ be an arbitrary element of the Lie algebra.
To write the Wilson loop using the color direction fields, and to reformulate the $G(2)$ Yang-Mills theory, we have to decompose $\mathscr{F}$ into the part $\mathscr{F}_{\tilde{H}}$ belonging to $\tilde{h}={\rm Lie}(\tilde{H})$, and the remaining part $\mathscr{F}_{G/\tilde{H}}$ using its commutators with $H_k$.
This is achieved by using double commutators in the case of $SU(N)$, see \cite{KKSS15}.  But, in the case of $G(2)$, we have to use sextuple commutators.
Its proof is given in Appendix \ref{sec:a}.
In this section, we give the explicit form of such  a decomposition for any representation.

\subsection{Decomposing SU(3)}

Before proceeding to the $G(2)$ case, we reconsider the $SU(3)$ case from the viewpoint of this paper.
For $SU(3)$, it is known \cite{KKSS15} that the maximal stability subgroup is $U(2)$ or $U(1)\times U(1)$.
In the case of the maximal stability subgroup $U(2)$ with generators $H_1$, $H_2$, $E_{\alpha^{(2)}}$ and $E_{-\alpha^{(2)}}$, the decomposition formula is written as
\begin{align}
	\mathscr{F}_{G/\tilde{H}} = \frac43 [H_2,[H_2,\mathscr{F}]] ,
\end{align}
while in the case of the maximal stability subgroup  $U(1)\times U(1)$, the decomposition formula is written as
\begin{align}
	\mathscr{F}_{G/\tilde{H}} = \sum_{j=1,2} [H_j,[H_j,\mathscr{F}]].
\end{align}
The derivation of these formulas is written in Appendix C and D in \cite{KKSS15}.
We rederive them using another method which can be applied also to  $G(2)$.
First, we consider the commutator of an arbitrary element of the Cartan subalgebra with $\mathscr{F}$.
An arbitrary element of the Cartan subalgebra can be written as $\nu\cdot H$, where $\nu$ is an arbitrary 2-dimensional vector.
Using the Cartan decomposition,
\begin{align}
	\mathscr{F} = \sum_{j=1,2} \mathscr{F}_j H_j + \sum_{\alpha\in\mathcal{R}_+}(\mathscr{F}_\alpha^* E_\alpha + \mathscr{F}_\alpha E_{-\alpha}),
\end{align}
and the commutation relation (\ref{3}), we can write the commutator as
\begin{align}
	[\nu\cdot H, \mathscr{F}] = \sum_{\alpha\in\mathcal{R}_+}(\nu\cdot\alpha)(\mathscr{F}_\alpha^* E_\alpha - \mathscr{F}_\alpha E_{-\alpha}).
\end{align}
Here, we choose $\gamma^1:=(\sqrt3/2,1/2)$ as $\nu$, which is orthogonal to $\alpha^{(1)}$, i.e., $\gamma^1\cdot\alpha^{(1)}=0$. Then the commutator reads
\begin{align}
	[\gamma^1\cdot H, \mathscr{F}] &= (\gamma^1\cdot\alpha^{(2)})(\mathscr{F}_{\alpha^{(2)}}^* E_{\alpha^{(2)}} - \mathscr{F}_{\alpha^{(2)}} E_{-\alpha^{(2)}}) + (\gamma^1\cdot\alpha^{(3)})(\mathscr{F}_{\alpha^{(3)}}^* E_{\alpha^{(3)}} - \mathscr{F}_{\alpha^{(3)}} E_{-\alpha^{(3)}}),
\end{align}
where the terms corresponding to $\alpha^{(1)}$ disappear.
By taking the commutator once more, we can eliminate another term.
For the vector $\gamma^2 := (0,1)$, which is orthogonal to $\alpha^{(2)}$, i.e., $\gamma^2\cdot\alpha^{(2)}=0$, the double commutator is written as
\begin{align}
	[\gamma^2\cdot H,[\gamma^1\cdot H, \mathscr{F}]] &= (\gamma^2\cdot\alpha^{(3)})(\gamma^1\cdot\alpha^{(3)})(\mathscr{F}_{\alpha^{(3)}}^* E_{\alpha^{(3)}} + \mathscr{F}_{\alpha^{(3)}} E_{-\alpha^{(3)}})\notag\\
	&= \frac34(\mathscr{F}_{\alpha^{(3)}}^* E_{\alpha^{(3)}} + \mathscr{F}_{\alpha^{(3)}} E_{-\alpha^{(3)}}).
\end{align}
Thus we obtain
\begin{align}
\label{24}
	 \mathscr{F}_{\alpha^{(3)}}^* E_{\alpha^{(3)}} + \mathscr{F}_{\alpha^{(3)}} E_{-\alpha^{(3)}}  = \frac43[\gamma^2\cdot H,[\gamma^1\cdot H, \mathscr{F}]].
\end{align}
In this way, we can extract the element of $\mathscr{F}$ corresponding to a particular positive root by taking the double commutator.  
For the other positive roots, the similar identity holds:
\begin{align}
\label{25}
	 \mathscr{F}_{\alpha^{(1)}}^* E_{\alpha^{(1)}} + \mathscr{F}_{\alpha^{(1)}} E_{-\alpha^{(1)}}  &= -\frac43[\gamma^3\cdot H,[\gamma^2\cdot H, \mathscr{F}]],\\
\label{26}
	 \mathscr{F}_{\alpha^{(2)}}^* E_{\alpha^{(2)}} + \mathscr{F}_{\alpha^{(2)}} E_{-\alpha^{(2)}} &= \frac43[\gamma^1\cdot H,[\gamma^3\cdot H, \mathscr{F}]],
\end{align}
where we have introduced $\gamma^3:=(\sqrt3/2,-1/2)$, which is orthogonal to $\alpha^{(3)}$.
Using these expressions, we can write the decomposition formula for any case of the maximal stability subgroup.
For $\tilde{H}= U(2)$, the decomposition formula is written as
\begin{align}
	\mathscr{F}_{\tilde{H}} &= \sum_{j_1,2}(\mathscr{F},H_j)H_j + (\mathscr{F}_{\alpha^{(2)}}^* E_{\alpha^{(2)}} + \mathscr{F}_{\alpha^{(2)}} E_{-\alpha^{(2)}}) \notag\\
	&=\sum_{j_1,2}(\mathscr{F},H_j)H_j +  [H_1,[H_1,\mathscr{F}]] - \frac13 [H_2,[H_2,\mathscr{F}]] \notag\\
	\mathscr{F}_{G/\tilde{H}} &= (\mathscr{F}_{\alpha^{(1)}}^* E_{\alpha^{(1)}} + \mathscr{F}_{\alpha^{(1)}} E_{-\alpha^{(1)}})
	+ (\mathscr{F}_{\alpha^{(3)}}^* E_{\alpha^{(3)}} + \mathscr{F}_{\alpha^{(3)}} E_{-\alpha^{(3)}}) 
\notag\\
	&= \frac43 [H_2,[H_2,\mathscr{F}]] ,
\end{align}
while for $\tilde{H}= U(1)\times U(1)$, the decomposition formula is written as
\begin{align}
	\mathscr{F}_{\tilde{H}} &= \sum_{j_1,2}(\mathscr{F},H_j)H_j \notag\\
	\mathscr{F}_{G/\tilde{H}} &=
	(\mathscr{F}_{\alpha^{(1)}}^* E_{\alpha^{(1)}} + \mathscr{F}_{\alpha^{(1)}} E_{-\alpha^{(1)}})
	+(\mathscr{F}_{\alpha^{(2)}}^* E_{\alpha^{(2)}} + \mathscr{F}_{\alpha^{(2)}} E_{-\alpha^{(2)}})
	+(\mathscr{F}_{\alpha^{(3)}}^* E_{\alpha^{(3)}} + \mathscr{F}_{\alpha^{(3)}} E_{-\alpha^{(3)}}) 
\notag\\
	&= \sum_{j=1,2} [H_j,[H_j,\mathscr{F}]],
\end{align}
where we have used the commuting property: 
\begin{align}
\label{18a}
[H_j,[H_k,\mathscr{F}]] = [[H_j,H_k],\mathscr{F}]+[H_k,[H_j,\mathscr{F}]]=[H_k,[H_j,\mathscr{F}]],
\end{align}
following from $[H_j,H_k] =0$.

\subsection{Decomposing G(2)}

Now we consider the $G(2)$ case.
We want to write the coset part $\mathscr{F}_{G/\tilde{H}}$ of $\mathscr{F}$ as a linear combination of multiple commutators with the Cartan generators:
\begin{align}
	\label{17}
	\mathscr{F}_{G/\tilde H}= \sum_{j_1,\dots,j_n\in\{1,2\}}\eta_{j_1\cdots j_n}[H_{j_1}, \cdots,[H_{j_n},\mathscr{F}]\cdots],
\end{align}
where the sum is over independent terms by taking account of the commuting property (\ref{18a}).
We can obtain (\ref{17}) for any choice of $\tilde{h}$ if, for every positive root $\beta$, the relevant shift part $\mathscr{R}_{\beta}$ is written in the form:
\begin{align}
\label{18}
	\mathscr{R}_\beta:= \mathscr{F}_\beta^* E_\beta + \mathscr{F}_\beta E_{-\beta}
	&= \sum_{j_1,\dots,j_n}\tilde{\eta}_{j_1\cdots j_n}[H_{j_1},\cdots,[H_{j_n},\mathscr{F}]\cdots].
\end{align}
In the following, we give a derivation of this fact (\ref{18}).
In the way similar to that written in the above for $SU(3)$, indeed, we can obtain (\ref{18}).
By using the commutation relation (\ref{3}), the commutator is calculated as 
\begin{align}
	[\nu\cdot H,\mathscr{F}] =\sum_{\alpha\in\mathcal{R}_+} (\nu\cdot \alpha)(\mathscr{F}_\alpha^*E_{\alpha}- \mathscr{F}_\alpha E_{-\alpha}).
\end{align}
If $\nu$ is chosen to be orthogonal to a particular $\alpha$, then the corresponding terms of $E_{\alpha}$ and $E_{-\alpha}$ disappear from  this expression.
Thus, by taking the commutator repeatedly, we can eliminate all shift terms except one   shift term $\mathscr{R}_{\beta}$ that corresponds to a particular positive root $\beta$.
Since there are six positive roots in $G(2)$, 
we have to eliminate five shift terms.
We can do so using the quintuple commutator:
\begin{align}
	\label{19}
	[\nu^1\cdot H,[\nu^2\cdot H,[\nu^3\cdot H,[\nu^4\cdot H,[\nu^5\cdot H,\mathscr{F}]]]]] =(\nu^1\cdot\beta)(\nu^2\cdot\beta)(\nu^3\cdot\beta)(\nu^4\cdot\beta)(\nu^5\cdot\beta)(\mathscr{F}_\beta^*E_{\beta}- \mathscr{F}_\beta E_{-\beta}),
\end{align}
where $\nu^1,\dots,\nu^5$ are appropriate 2-dimensional vectors.%
\footnote{
The reason why we need quintuple commutator is that we can eliminate only one term by taking the commutator once.
This is because 
the roots of $G(2)$ is 2-dimensional.
If the dimension of the root vectors is larger, we can eliminate more than one term by taking the commutator once. 
}
In this expression, the sign of the term of $E_\beta$ is opposite to that of $E_{-\beta}$.
To make both signs equal, we need to take the commutator once more.
We choose an  2-dimensional vector $\nu$ which is   non-orthogonal to $\beta$ to obtain the non-vanishing commutator of $\nu\cdot H$ and (\ref{19}):
\begin{align}
	[\nu\cdot H, (\ref{19})] = (\nu\cdot\beta)(\nu^1\cdot\beta)(\nu^2\cdot\beta)(\nu^3\cdot\beta)(\nu^4\cdot\beta)(\nu^5\cdot\beta)(\mathscr{F}_\beta^*E_{\beta}+ \mathscr{F}_\beta E_{-\beta}).
\end{align}
Thus we obtain the key relation:
\begin{align}
	&\mathscr{R}_\beta = \frac1N[\nu\cdot H,[\nu^1\cdot H,[\nu^2\cdot H,[\nu^3\cdot H,[\nu^4\cdot H,[\nu^5\cdot H,\mathscr{F}]]]]]], \notag\\
	&N :=(\nu\cdot\beta)(\nu^1\cdot\beta)(\nu^2\cdot\beta)(\nu^3\cdot\beta)(\nu^4\cdot\beta)(\nu^5\cdot\beta). 
	\label{21}
\end{align}
Although this expression is nothing but the desired one (\ref{18}), it should be remarked that the coefficients $\tilde{\eta}_{j_1\cdots j_6}$ is not uniquely determined.
If we multiply $\nu$ by a constant, the coefficients $\tilde{\eta}_{j_1\cdots j_6}$ do not change.
This point will be observed more concretely shortly. 

\begin{figure}[t]
\begin{center}
\includegraphics[width=0.2\hsize]{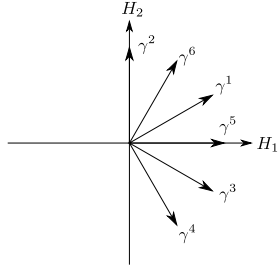}
\caption{Unit vectors which is orthogonal to one of the positive roots.}
\label{fig:orvec}
\end{center}
\end{figure}

To obtain the expression (\ref{21}) for each positive root concretely, we introduce six unit vectors $\gamma^a$ ($a=1,\dots,6$) such that $\gamma^a$ is positive and orthogonal to one of the positive roots, say $\alpha^{(a)}$ ($a=1,\dots,6$):
\begin{align}
	&\gamma^1=(\frac{\sqrt3}2,\frac12),\;\gamma^2=(0,1),\;\gamma^3=(\frac{\sqrt3}2,-\frac12),\;\gamma^4=(\frac12,-\frac{\sqrt3}2),\;\gamma^5=(1,0),\; \gamma^6=(\frac12,\frac{\sqrt3}2).
\end{align}
See FIG.~\ref{fig:orvec}. 
Consequently these vectors satisfy the following conditions:
\begin{align}
	\gamma^a \bot \alpha^{(a)},\; 
\gamma^4\parallel \alpha^{(1)},\;
\gamma^5\parallel \alpha^{(2)},\;
\gamma^6\parallel \alpha^{(3)},\;
\gamma^1\parallel \alpha^{(4)},\;
\gamma^2\parallel \alpha^{(5)},\;
\gamma^3\parallel \alpha^{(6)}.
\end{align}

For example, $\mathscr{R}_{\alpha^{(1)}}$ is obtained as
\begin{align}
	&\mathscr{R}_{\alpha^{(1)}} = N^{-1} [\gamma\cdot H,[\gamma^2\cdot H,[\gamma^3\cdot H,[\gamma^4\cdot H,[\gamma^5\cdot H,[\gamma^6\cdot H,\mathscr{F}]]]]]]  ,
\notag\\
	&N = (\gamma\cdot\alpha^{(1)})(\gamma^2\cdot\alpha^{(1)})(\gamma^3\cdot\alpha^{(1)})(\gamma^4\cdot\alpha^{(1)})(\gamma^5\cdot\alpha^{(1)})(\gamma^6\cdot\alpha^{(1)}) ,
\end{align}
where $\gamma$ is an arbitrary 2-dimensional vector that is not orthogonal to $\alpha^{(1)}$.

To obtain more explicit form, we put an arbitrary 2-dimensional vector $\gamma$ in the form:
$\gamma= a\gamma^1+b\gamma^4$ (orthogonal decomposition of $\gamma$) 
where
$\gamma^1$ is orthogonal to $\alpha^{(1)}$ and $\gamma^4$ is parallel to $\alpha^{(1)}$:
$\gamma^1 \bot \alpha^{(1)}$
and
$\gamma^4\parallel \alpha^{(1)}$.
Here  
$b\neq0$ to avoid $\gamma\cdot\alpha^{(1)}\equiv b\gamma^4\cdot \alpha^{(1)}=0$. 
Using $\gamma^1\cdot \alpha^{(1)}=0$, we have 
$\gamma^2\cdot \alpha^{(1)}=-\sqrt3|\alpha^{(1)}|/2$, $\gamma^3\cdot\alpha^{(1)}=\sqrt3|\alpha^{(1)}|/2$, $\gamma^4\cdot\alpha^{(1)}=|\alpha^{(1)}|$, $\gamma^5\cdot\alpha^{(1)}=|\alpha^{(1)}|/2$, $\gamma^6 \cdot \alpha^{(1)} = -|\alpha^{(1)}|/2$,
and 
$\gamma\cdot\alpha^{(1)}=b\gamma^4\cdot \alpha^{(1)}=b |\alpha^{(1)}|$, which yields 
\begin{align}
	N = b\frac3{16}  |\alpha^{(1)}|^6=\frac{3b}{16} . 
\end{align}
Combining the result $N^{-1} = \frac{16}{3b}$ with 
$
	 \gamma\cdot H = a \gamma^1\cdot H + b \gamma^4 \cdot H  
$,
therefore, we obtain
\begin{align}
	\mathscr{R}_{\alpha^{(1)}} &=\frac{16}3[\gamma^4\cdot H,[\gamma^2\cdot H,[\gamma^3\cdot H,[\gamma^4\cdot H,[\gamma^5\cdot H,[\gamma^6\cdot H,\mathscr{F}]]]]]] \notag\\
	&\quad + c_1[\gamma^1\cdot H,[\gamma^2\cdot H,[\gamma^3\cdot H,[\gamma^4\cdot H,[\gamma^5\cdot H,[\gamma^6\cdot H,\mathscr{F}]]]]]], 
\end{align}
where $c_1=16a/3b$.
We can see from this expression that the non-uniqueness of an expression of $\mathscr{R}_{\alpha^{(1)}}$ comes from the fact that the following sextuple commutator is identically vanishing:
\begin{align}
	&\mathcal{Z} := [\gamma^1\cdot H,[\gamma^2\cdot H,[\gamma^3\cdot H,[\gamma^4\cdot H,[\gamma^5\cdot H,[\gamma^6\cdot H,\mathscr{F}]]]]]] 
\notag \\
 &\quad= \sum_{j_1,\dots,j_6\in{1,2}}\bar{\zeta}_{j_1\cdots j_6}[H_{j_1},[H_{j_2},[H_{j_3},[H_{j_4},[H_{j_5},[H_{j_6},\mathscr{F}]]]]]]= 0, 
\notag\\
	&\bar{\zeta}_{11112} = \frac3{16},\; \bar{\zeta}_{111222}=-\frac58, \; \bar{\zeta}_{122222}=\frac3{16} .
\end{align}
Thus, the non-uniqueness of the decomposition formula is attributed to degree of freedom  due to one parameter $c_1$.

In the same way as the above, we obtain
\begin{align}
	\mathscr{R}_{\alpha^{(2)}} &=\frac{16}3[\gamma^5\cdot H,[\gamma^1\cdot H,[\gamma^3\cdot H,[\gamma^4\cdot H,[\gamma^5\cdot H,[\gamma^6\cdot H,\mathscr{F}]]]]]] + c_2 \mathcal{Z}, \\
	\mathscr{R}_{\alpha^{(3)}} &=-\frac{16}3[\gamma^6\cdot H,[\gamma^1\cdot H,[\gamma^2\cdot H,[\gamma^4\cdot H,[\gamma^5\cdot H,[\gamma^6\cdot H,\mathscr{F}]]]]]] + c_3 \mathcal{Z}, \\
	\mathscr{R}_{\alpha^{(4)}} &=\frac{16}3(\sqrt{3})^6[\gamma^1\cdot H,[\gamma^1\cdot H,[\gamma^2\cdot H,[\gamma^3\cdot H,[\gamma^5\cdot H,[\gamma^6\cdot H,\mathscr{F}]]]]]]+ c_4 \mathcal{Z}, \\
	\mathscr{R}_{\alpha^{(5)}} &=\frac{16}3(\sqrt3)^6[\gamma^2\cdot H,[\gamma^1\cdot H,[\gamma^2\cdot H,[\gamma^3\cdot H,[\gamma^4\cdot H,[\gamma^6\cdot H,\mathscr{F}]]]]]] + c_5\mathcal{Z}, \\
	\mathscr{R}_{\alpha^{(6)}} &=-\frac{16}3(\sqrt3)^6[\gamma^3\cdot H,[\gamma^1\cdot H,[\gamma^2\cdot H,[\gamma^3\cdot H,[\gamma^4\cdot H,[\gamma^5\cdot H,\mathscr{F}]]]]]] + c_6 \mathcal{Z}, 
\end{align}	
where $c_2,\dots,c_6$ are arbitrary constants.
Here, we have used the commuting property (\ref{18a}).

Collecting an appropriate set of $\mathscr{R}_\alpha$, we obtain the desired decomposition formula corresponding to each category of representations given in (\ref{12})(\ref{13})(\ref{14}):
\begin{enumerate}
\item 
For $\tilde{H} = U(2)\subset SU(3)$ with the generators $H_1$, $H_2$, $E_{\alpha^{(1)}}$ and $E_{-\alpha^{(1)}}$, the highest weight is $\Lambda= (m/2, m/(2\sqrt3))$.
Here, we redefine the highest weight as $(0,m/\sqrt3)$ in order to obtain simpler form.
This is possible because new one is obtained by  acting a Weyl group element on an old one.
Thus $\tilde{H}$ has the generators $H_1$, $H_2$, $E_{\alpha^{(2)}}$ and $E_{-\alpha^{(2)}}$.
The 
$\tilde{H}$-commutative part $\mathscr{F}_{\tilde{H}}$ and the 
coset part $\mathscr{F}_{G/\tilde{H}}$ of $\mathscr{F}$ are given by
\begin{align}
 \mathscr{F}_{\tilde{H}} =& \sum_{j=1,2} (\mathscr{F},H_j)H_j   + \mathscr{R}_{\alpha^{(2)}} , 
\nonumber\\
	\mathscr{F}_{G/\tilde{H}} =& \mathscr{R}_{\alpha^{(1)}} + \mathscr{R}_{\alpha^{(3)}} + \mathscr{R}_{\alpha^{(4)}} + \mathscr{R}_{\alpha^{(5)}} + \mathscr{R}_{\alpha^{(6)}}. 
\end{align}
The explicit form is given by
\begin{align}
	\mathscr{F}_{\tilde{H}} &= \sum_{j=1,2}(\mathscr F, H_j)H_j + \sum_{j_1,\dots,j_6}\tilde{\zeta}_{j_1\cdots j_6}[H_{j_1},[H_{j_2},[H_{j_3},[H_{j_4},[H_{j_5},[H_{j_6},\mathscr{F}]]]]]] + c_2\mathcal{Z}, \notag\\
	\tilde{\zeta}_{111111} &= 1,\; \tilde{\zeta}_{111122} = -\frac{10}3,\; \tilde{\zeta}_{112222} = 1,\notag\\ 
	\mathscr{F}_{G/\tilde{H}} &= \sum_{j_1,\dots,j_6}\zeta_{j_1\cdots j_6}[H_{j_1},[H_{j_2},[H_{j_3},[H_{j_4},[H_{j_5},[H_{j_6},\mathscr{F}]]]]]] + c\mathcal{Z} ,
\notag\\
	\zeta_{111122} &= \frac{721}3,\; \zeta_{112222} = -154,\; \zeta_{222222}=27, 
	\label{zeta}
\end{align}
where the other $\zeta_{j_1\cdots j_6}$s and $\tilde \zeta_{j_1\cdots j_6}$ are zero and $c := c_1+ c_3+ c_4+ c_5+ c_6$.
The simplest choice is $c=0$ and $c_2=0$.

\item 
For $\tilde{H}=U'(2) (\not\subset SU(3))$ with the generators $H_1$, $H_2$, $E_{\alpha^{(5)}}$ and $E_{-\alpha^{(5)}}$, 
\begin{align}
 \mathscr{F}_{\tilde{H}} =& \sum_{j=1,2} (\mathscr{F},H_j)H_j   + \mathscr{R}_{\alpha^{(5)}} , 
\nonumber\\
	\mathscr{F}_{G/\tilde{H}} =& \mathscr{R}_{\alpha^{(1)}} + \mathscr{R}_{\alpha^{(2)}} + \mathscr{R}_{\alpha^{(3)}} + \mathscr{R}_{\alpha^{(4)}} + \mathscr{R}_{\alpha^{(6)}}. 
\end{align}
We obtain
\begin{align}
	\mathscr{F}_{\tilde{H}} &= \sum_{j=1,2}(\mathscr F, H_j)H_j +  \sum_{j_1,\dots,j_6}\tilde{\zeta}'_{j_1\cdots j_6}[H_{j_1},[H_{j_2},[H_{j_3},[H_{j_4},[H_{j_5},[H_{j_6},\mathscr{F}]]]]]] + c_5\mathcal{Z}, \notag\\
	\tilde{\zeta}'_{111122} &= 27,\; \tilde{\zeta}'_{112222} = -90,\; \tilde{\zeta}'_{222222} = 27,\notag\\ 
	\mathscr{F}_{G/\tilde{H}} &= \sum_{j_1,\dots,j_6}\zeta'_{j_1\cdots j_6}[H_{j_1},[H_{j_2},[H_{j_3},[H_{j_4},[H_{j_5},[H_{j_6},\mathscr{F}]]]]]] + c'\mathcal{Z} ,
\notag\\
	\zeta'_{111111} &= 1,\; \zeta'_{111122} =210, \; \zeta'_{112222}=-63, 
	\label{zeta'}
\end{align}
where the other $\zeta'_{j_1\cdots j_6}$s and $\tilde \zeta_{j_1\cdots j_6}$ are zero.
We can take the simplest choice $c'=0$ and $c_5=0$.

\item 
For $\tilde{H} = U(1)\times U(1)$,
\begin{align}
 \mathscr{F}_{\tilde{H}} =& \sum_{j=1,2} (\mathscr{F},H_j)H_j   , 
\nonumber\\
	\mathscr{F}_{G/\tilde{H}} =& \mathscr{R}_{\alpha^{(1)}} + \mathscr{R}_{\alpha^{(2)}} + \mathscr{R}_{\alpha^{(3)}} + \mathscr{R}_{\alpha^{(4)}} +\mathscr{R}_{\alpha^{(5)}}+  \mathscr{R}_{\alpha^{(6)}}. 
\end{align}
We obtain
\begin{align}
	\mathscr{F}_{G/\tilde{H}} &= \sum_{j_1,\dots,j_6}\zeta''_{j_1\cdots j_6}[H_{j_1},[H_{j_2},[H_{j_3},[H_{j_4},[H_{j_5},[H_{j_6},\mathscr{F}]]]]]] + c''\mathcal{Z} ,
\notag\\
	\zeta''_{111111} &= 1,\; \zeta''_{111122} =237, \; \zeta''_{112222}=-153, \; \zeta''_{222222}=27, 
	\label{zeta''}
\end{align}
where the other $\zeta''_{j_1\cdots j_6}$s are zero.
We can take the simplest choice $c''=0$.

\end{enumerate}

Using the decomposition formula, we can define the field decomposition in the similar way to the case of the gauge group $SU(N)$.
For this purpose, we define the color direction field for $G(2)$ as
\begin{align}
	\bm n_j(x) := {\rm Ad}_{g(x)}(H_j),
\end{align}
where ${\rm Ad}_{g(x)}$ is the adjoint representation of $g(x)$, where $g(x)$ is an arbitrary group-valued field.\footnote{
This definition of the color direction field is consistent with the definition adopted in the previous works for the gauge group $SU(N)$ because
\begin{align}
	R(\bm{n}_j(x)) = R({\rm Ad}_{g(x)}(H_j)) = R(g(x))R(H_j)R(g(x))^\dag. \notag
\end{align}
Here we have used the same notation $R$ to denote the group representation and the corresponding algebra representation, which does not cause the confusion because the domains are different.}
For any Lie algebra valued field $\mathscr F(x)$, by applying the decomposition formula to ${\rm Ad}_{g^{-1}(x)}(\mathscr F(x))$ and operating ${\rm Ad}_{g(x)}$ on the both sides, we can decompose $\mathscr F(x)$ into the part $\mathscr F_{\tilde H}(x)$ belonging to ${\rm Ad}_{g(x)}(\tilde h)$ and the remaining part $\mathscr F_{G/\tilde H}(x)$: 
\begin{align}
\mathscr{F}(x) =&	\mathscr{F}_{\tilde{H}}(x) + \mathscr{F}_{G/\tilde{H}}(x) ,
\notag\\
 &\mathscr{F}_{\tilde{H}}(x) = \sum_{j=1,2} (\mathscr{F}(x),\bm{n}_{j}(x))\bm{n}_{j}(x) +  \sum_{j_1,\dots,j_6} \xi_{j_1\dots j_6} [\bm{n}_{j_1}(x),\cdots,[\bm{n}_{j_5}(x),[\bm{n}_{j_6}(x),\mathscr{F}(x)]]\cdots] , 
\notag\\
	&	\mathscr{F}_{G/\tilde{H}}(x) = \sum_{j_1,\dots,j_6}\eta_{j_1\cdots j_6}[\bm{n}_{j_1}(x),\cdots,[\bm{n}_{j_5}(x),[\bm{n}_{j_6}(x),\mathscr{F}(x)]]\cdots] ,
	\label{F-decomp}
\end{align}
where $\xi_{j_1\cdots j_6}$ and $\eta_{j_1\cdots j_6}$ are appropriate coefficients specified by the maximal stability subgroup.
We decompose the Yang-Mills field $\mathscr{A}_\mu(x)$ into two pieces, $\mathscr{V}_\mu(x)$ and $\mathscr{X}_\mu(x)$:
\begin{align}
	\mathscr{A}_\mu(x) = \mathscr{V}_\mu(x) + \mathscr{X}_\mu(x)  ,
\end{align}
where the decomposed fields $\mathscr{V}_\mu(x)$ and $\mathscr{X}_\mu(x)$  are obtained as the solution of the defining equations:
\begin{align}
	&0 = \mathscr{D}_\mu[\mathscr{V}]\bm{n}_j(x) := \partial_\mu\bm{n}_j(x) - ig_{{}_{\rm YM}}[\mathscr{V}_\mu(x), \bm{n}_j(x)] ,
	\label{def-1}
\\
	&0 = \mathscr{X}_\mu(x)_{\tilde{H}} \Leftrightarrow \mathscr{X}_\mu(x) = \sum_{j_1,\dots,j_6}\eta_{j_1\cdots j_6}[\bm{n}_{j_1}(x),\cdots,[\bm{n}_{j_5}(x),[\bm{n}_{j_6}(x),\mathscr{X}_\mu(x)]]\cdots].
	\label{def-2}
\end{align}
Using the first defining equation (\ref{def-1}), we find
\begin{align}
 \mathscr{D}_\mu[\mathscr{A}]\bm{n}_j(x) 
 =& \mathscr{D}_\mu[\mathscr{V}]\bm{n}_j(x)  - ig_{{}_{\rm YM}}[\mathscr{X}_\mu(x), \bm{n}_j(x)]
 =  ig_{{}_{\rm YM}}[ \bm{n}_j(x), \mathscr{X}_\mu(x)]
 . 
\end{align}
By substituting this relation into the second defining equation (\ref{def-2}),  $\mathscr{X}_\mu(x)$ is rewritten   as
\begin{align}
 \mathscr{X}_\mu(x) 
=& - ig_{{}_{\rm YM}}^{-1} \sum_{j_1,\dots,j_6}\eta_{j_1\cdots j_6}[\bm{n}_{j_1}(x),\cdots,[\bm{n}_{j_5}(x),  \mathscr{D}_\mu[\mathscr{A}]\bm{n}_{j_6}(x) ]\cdots] 
 .
\end{align}
Then $\mathscr{V}_\mu(x)$ is written as 
\begin{align}
 \mathscr{V}_\mu(x) 
=& \mathscr{A}_\mu(x) - \mathscr{X}_\mu(x) 
\nonumber\\
=& \mathscr{A}_\mu(x) +  ig_{{}_{\rm YM}}^{-1} \sum_{j_1,\dots,j_6}\eta_{j_1\cdots j_6}[\bm{n}_{j_1}(x),\cdots,[\bm{n}_{j_5}(x),  \mathscr{D}_\mu[\mathscr{A}]\bm{n}_{j_6}(x) ]\cdots] 
 .
\end{align}
Thus $\mathscr{V}_\mu(x)$ and $\mathscr{X}_\mu(x)$ are written in terms of the original Yang-Mills field $\mathscr{A}_\mu(x)$ and the color fields $\bm{n}_j(x)$.

Notice that $\mathscr{V}_\mu(x)$ is further cast into 
\begin{align}
 \mathscr{V}_\mu(x) 
=& \mathscr{A}_\mu(x) 
+  \sum_{j_1,\dots,j_6}\eta_{j_1\cdots j_6}[\bm{n}_{j_1}(x),\cdots,[\bm{n}_{j_5}(x),  [\mathscr{A}_\mu(x),\bm{n}_{j_6}(x)] ]\cdots] 
\nonumber\\&
+  ig_{{}_{\rm YM}}^{-1} \sum_{j_1,\dots,j_6}\eta_{j_1\cdots j_6}[\bm{n}_{j_1}(x),\cdots,[\bm{n}_{j_5}(x),  \partial_\mu \bm{n}_{j_6}(x) ]\cdots] 
\nonumber\\
=& \sum_{j=1,2} (\mathscr{A}_\mu(x),\bm{n}_{j}(x))\bm{n}_{j}(x) +  \sum_{j_1,\dots,j_6} \xi_{j_1\dots j_6} [\bm{n}_{j_1}(x),\cdots,[\bm{n}_{j_5}(x),[\bm{n}_{j_6}(x),\mathscr{A}_\mu(x)]]\cdots]
\nonumber\\&
 +  ig_{{}_{\rm YM}}^{-1} \sum_{j_1,\dots,j_6}\eta_{j_1\cdots j_6}[\bm{n}_{j_1}(x),\cdots,[\bm{n}_{j_5}(x),  \partial_\mu \bm{n}_{j_6}(x) ]\cdots] 
 ,
\end{align}
where we have applied the the formula (\ref{F-decomp}) to $\mathscr{A}_\mu(x)$ in the last step. 
Therefore, $\mathscr{V}_\mu(x)$ is decomposed into $\mathscr{C}_\mu(x)$ and $\mathscr{B}_\mu(x)$:
\begin{align}
 \mathscr{V}_\mu(x) 
=& \mathscr{C}_\mu(x) + \mathscr{B}_\mu(x) 
\nonumber\\
 \mathscr{C}_\mu(x) :=& \sum_{j=1,2} (\mathscr{A}_\mu(x),\bm{n}_{j}(x))\bm{n}_{j}(x) +  \sum_{j_1,\dots,j_6} \xi_{j_1\dots j_6} [\bm{n}_{j_1}(x),\cdots,[\bm{n}_{j_5}(x),[\bm{n}_{j_6}(x),\mathscr{A}_\mu(x)]]\cdots] ,
\nonumber\\ 
 \mathscr{B}_\mu(x) :=&  ig_{{}_{\rm YM}}^{-1} \sum_{j_1,\dots,j_6}\eta_{j_1\cdots j_6}[\bm{n}_{j_1}(x),\cdots,[\bm{n}_{j_5}(x),  \partial_\mu \bm{n}_{j_6}(x) ]\cdots] \label{decom_cb}
 .
\end{align}
For the sake of convenience, we define the field $\bm{m}(x)$ for the highest weight $\Lambda=(\Lambda_1, \Lambda_2)$ by
\begin{align}
\bm{m}(x) : = \Lambda_j \bm{n}_j(x) .
\end{align}
Here $\mathscr{C}_\mu(x)$ commutes with 
$\bm{m}(x)$:
\begin{align}
[ \mathscr{C}_\mu(x), \bm{m}(x) ] = 0 \label{cm}
 ,
\end{align}
while $\mathscr{B}_\mu(x)$ is orghogonal to $\bm{n}_j(x)$:
\begin{align}
 (\mathscr{B}_\mu(x),\bm{n}_{j}(x)) = 0
 .
\end{align}
The first term in the right-hand side of $\mathscr{C}_\mu(x)$ corresponds to the element of the Cartan subalgebra ${\rm Lie}(H)$ and the second term to the remaining part ${\rm Lie}(\tilde{H})-{\rm Lie}(H)$ which vanishes when the maximal stability group coincides with the maximal torus group $\tilde{H}=H$ (This is the case for the maximal option of $SU(N)$). Notice that   
$\mathscr{B}_\mu(x)$ is the extension of the $SU(N)$ Cho connection to $G(2)$. 
An appropriate set of the above fields will be used in the reformulation of the $G(2)$ Yang-Mills theory.

We suppose that the dominant mode for quark confinement is  the restricted field  $\mathscr{V}_\mu(x)$ extracted from the original $G(2)$ Yang-Mills field $\mathscr{A}_\mu(x)$ through the decomposition given in the above. 
In fact, this observation is exemplified for the $G(2)$ Wilson loop operator by using the non-Abelian Stokes theorem in the same manner as in $SU(N)$, as given in the next section. 


\section{Non-Abelian Stokes theorem}

In this section, we derive the non-Abelian Stokes theorem for the Wilson loop operator in an arbitrary representation of $G(2)$ gauge group using the color direction fields $\bm{n}_k$.

\subsection{General gauge group}

Before proceeding to the case of the gauge group $G(2)$, we discuss the general case.

It is known \cite{Kondo08,KKSS15,MK15} that the Wilson loop operator defined for any Lie algebra valued Yang-Mills field $\mathscr{A}$ and the irreducible (unitary) representation $R$ is cast into the following (path-integral) representation\footnote{
Strictly speaking, we should write $F^g_{\mu\nu}$ in (\ref{wl}) as
\begin{align}
	&F^g_{\mu\nu} = \kappa\{ \partial_\mu{\rm tr}\bigl(R(\bm m(x))R(\mathscr A_\nu(x))\bigr) 
	- \partial_\nu{\rm tr}\bigl(R(\bm m(x))R(\mathscr A_\mu(x))\bigr)
	+ig_{{}_{\rm YM}}{\rm tr}\bigl(R(\bm m(x))([\Omega_\mu(x),\Omega_\nu(x)])\bigr)\}, \notag\\
	&R(\bm m(x)) = \Lambda_j R(g(x))R(H_j)R(g(x))^\dag = \Lambda_j R({\rm Ad}_{g(x)}(H_j)), \notag\\
	&\Omega_\mu(x) = ig^{-1}_{{}_{\rm YM}}R(g(x))\partial_\mu R(g(x))^\dag. \notag
\end{align}
To simplify the notation, we omit the symbol $R(\cdot)$ throughout this section, Appendix \ref{sec:dnast} and \ref{sec:c}.}:
\begin{align}
	&W_C[\mathscr{A}] = \int[d\mu(g)]_\Sigma \exp \left[ -ig_{{}_{\rm YM}} \int_{\Sigma:\partial\Sigma=C} F^g\right], \notag\\
	&F^g = \frac12 F^g_{\mu\nu} dx^\mu\wedge dx^\nu \notag\\
	&F^g_{\mu\nu}(x) = \kappa\{ \partial_\mu {\rm tr}(\bm{m}(x)\mathscr{A}_\nu(x)) - \partial_\nu {\rm tr}(\bm{m}(x)\mathscr{A}_\mu(x)) + ig_{{}_{\rm YM}} {\rm tr}(\bm{m}(x) [\Omega_\mu(x),\Omega_\nu(x)])\}, \notag\\
	&\Omega_\mu(x) := ig_{{}_{\rm YM}}^{-1} g(x)\partial_\mu g^\dag(x), \quad g(x)\in G, \label{wl}
\end{align}
where $[d\mu(g)]_\Sigma$ is the product measure of the Haar measure on the gauge group $G$ over $\Sigma$ and $\Lambda$ in $\bm m:= \Lambda_j \bm n_j$ is the highest weight vector of the representation $R$.
Here the gauge-invariant field strength $F^g_{\mu\nu}$ is equal to the non-Abelian field strength $\mathscr{F}_{\mu\nu}[\mathscr{V}]:=\partial_\mu \mathscr{V}_\nu- \partial_\nu \mathscr{V}_\mu -ig_{{}_{\rm YM}} [\mathscr{V}_\mu, \mathscr{V}_\nu]$ of the restricted field $\mathscr{V}_\mu$ (in the decomposition $\mathscr{A}=\mathscr{V}+\mathscr{X}$) projected to the color field $\bm{m}$:
\begin{align}
F^g_{\mu\nu} =  {\rm tr}\{ \bm{m} \mathscr{F}_{\mu\nu} [\mathscr{V}]\} = \Lambda_j f_{\mu\nu}^{(j)}, \quad
 f_{\mu\nu}^{(j)} =  {\rm tr}\{ \bm{n}_j \mathscr{F}_{\mu\nu} [\mathscr{V}]\} . \label{v}
\end{align}
Therefore, the restricted field $\mathscr{V}_\mu$ is regarded as the dominant mode for quark confinement, since the remaining field $\mathscr{X}_\mu$ does not contribute to the Wilson loop operator. 
The derivation of this fact is given in Appendix \ref{sec:c}.

Let $\mathscr{F}$ be an arbitrary element of the Lie algebra $\mathcal{G}={\rm Lie}(G)$.
Suppose that $\mathscr{F}$ is decomposed as
\begin{align}
\mathscr{F} =&	\mathscr{F}_{\tilde{H}} + \mathscr{F}_{G/\tilde{H}} ,
\notag\\
&\mathscr{F}_{\tilde{H}} = \sum_{j=1}^r (\mathscr{F},H_j)H_j +  \sum_{j_1,\dots,j_n} \xi_{j_1\dots j_n}[H_{j_1},\cdots,[H_{j_n},\mathscr{F}]\cdots], 
\notag\\
	\label{40}
	&\mathscr{F}_{G/\tilde{H}} = \sum_{j_1,\dots,j_n} \eta_{j_1\cdots j_n}[H_{j_1},\cdots,[H_{j_n},\mathscr{F}]\cdots],
\end{align}
where $r$ is the rank of the gauge group.
At least, this relation for the decomposition has already been  proved for $G(2)$ 
in the previous section, and the method is applicable to any semi-simple compact Lie group.

In order to complete the non-Abelian Stokes theorem, 
we can follow the same procedures as those for $SU(N)$ given in \cite{MK15}, if $g^\dag\partial_\mu\bm{m}g$ does not have the part belonging to the ${\rm Lie}(\tilde H)$:
\begin{align}
	\label{42}
	(g^\dag\partial_\mu\bm{m}g)_{\tilde{H}} = 0.
\end{align}
This enables us to rewrite   $[\Omega_\mu(x),\Omega_\nu(x)]$ in terms of the color fields $\bm{n}_i(x):={\rm Ad}_{g(x)}(H_j)$, which is indeed  shown in Appendix \ref{sec:dnast}.

The relevant relation (\ref{42}) is indeed verified as follows. 
By applying (\ref{40}) to ${\rm Ad}_{g^{-1}}(\partial_\mu\bm{m})$, we obtain the decomposition:
\begin{align}
	(g^\dag\partial_\mu\bm mg)_{\tilde{H}} = \sum_{j=1}^r \kappa{\rm tr}(g^\dag\partial_\mu\bm{m}gH_j)H_j +  \sum_{i_1,\dots,i_n} \xi_{i_1\dots i_n}[H_{i_1},\cdots,[H_{i_n},g^\dag\partial_\mu\bm{m}g]\cdots]. 
	\label{43}
\end{align}
The first term on the right-hand side vanishes,
since
\begin{align}
	{\rm tr}(g^\dag\partial_\mu\bm{m}gH_j) &= \Lambda_i{\rm tr}(g^\dag\partial_\mu(gH_ig^\dag)gH_j) \notag\\
	&= \Lambda_i{\rm tr}(g^\dag\partial_\mu gH_iH_j + H_i\partial_\mu g^\dag gH_j) \notag\\
	&= \Lambda_i{\rm tr}(g^\dag\partial_\mu gH_iH_j - H_i g^\dag \partial_\mu gH_j) \notag\\
	&= \Lambda_i{\rm tr}(g^\dag\partial_\mu gH_iH_j - g^\dag \partial_\mu gH_jH_i) \notag\\
	&= \Lambda_i{\rm tr}(g^\dag\partial_\mu g[H_i,H_j])  \notag\\
	&=0,
\end{align}
where we have used $g^\dag g=1$ in the second equality, the relation $\partial_\mu g^\dag g= -g^\dag \partial_\mu g$ following from $\partial_\mu(gg^\dag)=0$ in the third equality and the cyclicity of the trace in the fourth equality.
In addition, by taking account of 
\begin{align}
	g^\dag\partial_\mu\bm{m}g &= g^\dag \partial_\mu(g\Lambda\cdot Hg^\dag)g \notag\\
	&= g^\dag\partial_\mu g\Lambda\cdot H + \Lambda\cdot H \partial_\mu g^\dag g \notag\\
	&= -\partial_\mu g^\dag g\Lambda\cdot H + \Lambda\cdot H \partial_\mu g^\dag g \notag\\
	&= [\Lambda\cdot H, \partial_\mu g^\dag g] ,
\end{align}
the second term is rewritten as
\begin{align}
	\sum_{i_1,\dots,i_n} \xi_{i_1\dots i_n}[H_{i_1},\cdots,[H_{i_n},g^\dag\partial_\mu\bm{m}g]\cdots] &= 
	\sum_{i_1,\dots,i_n} \xi_{i_1\dots i_n}[H_{i_1},\cdots,[H_{i_n},[\Lambda\cdot H, \partial_\mu g^\dag g]]\cdots] \notag\\
	&= \sum_{i_1,\dots,i_n} \xi_{i_1\dots i_n}[\Lambda\cdot H,[H_{i_1},\cdots,[H_{i_n}, \partial_\mu g^\dag g]\cdots]] \notag\\
	&= [\Lambda\cdot H, (\partial_\mu g^\dag g)_{\tilde{H}}] ,
	\label{46}
\end{align}
where we have used the commuting property (\ref{18a}) in the second equality and (\ref{40}) for $\mathscr{F}=\partial_\mu g^\dag g$ in the third equality.
By substituting the Cartan decomposition of $(\partial_\mu g^\dag g)_{\tilde{H}}$ in $\tilde{h}$ given by
\begin{align}
	 (\partial_\mu g^\dag g)_{\tilde{H}}= \sum_{j=1}^r \kappa{\rm tr}(\partial_\mu g^\dag gH_j)H_j   + \sum_{\alpha\in\mathcal{R}_+:E_{\pm\alpha}\in\tilde{h}}((\partial_\mu g^\dag g)_\alpha^* E_\alpha + (\partial_\mu g^\dag g)_\alpha E_{-\alpha}),
\end{align}
into  (\ref{46}), we find that the second term also vanishes,
\begin{align}
	(\ref{46})&= [\Lambda\cdot H, \sum_{\alpha\in \mathcal{R}_+:E_{\pm\alpha}\in \tilde{h}}((\partial_\mu g^\dag g)_\alpha^*E_\alpha + (\partial_\mu g^\dag g)_\alpha E_{-\alpha})] \notag\\
	&= \sum_{\alpha\in \mathcal{R}_+: E_{\pm\alpha}\in \tilde{h}}\Lambda\cdot\alpha((\partial_\mu g^\dag g)_\alpha^*E_\alpha - (\partial_\mu g^\dag g)_\alpha E_{-\alpha}) \notag\\
	&=0,
\end{align}
where we have used $\Lambda\cdot\alpha=0$ for $\alpha$ satisfying $E_\alpha \in \tilde{h}$.
Thus we have confirmed (\ref{42}).

Thus we obtain the final form of Wilson loop operator as\footnote{
We can rewrite $F^g_{\mu\nu}$ of (\ref{NAST-gen}) using the inner product instead of using the trace as
\begin{align}
	F^g_{\mu\nu} = \partial_\mu(\bm m,\mathscr A) - \partial_\nu(\bm m,\mathscr A) + ig_{{}_{\rm YM}}\sum_{i_1,\dots,i_n}\eta_{i_1\cdots i_n}(\bm m(x), [\partial_\mu \bm n_{i_1}(x),[\bm n_{i_2}(x),\cdots,[\bm n_{i_{n-1}}(x),\partial_\nu\bm n_{i_n}(x)]\cdots]]), \notag
\end{align}
so that the Wilson loop depends on the representation only through the highest weight vector $\Lambda$.}
\begin{align}
	&W_C[\mathscr{A}] = \int[d\mu(g)]_\Sigma \exp \left[ -ig_{{}_{\rm YM}} \int_{\Sigma:\partial\Sigma=C} F^g\right], \notag\\
	&F^g = \frac12 F^g_{\mu\nu} dx^\mu\wedge dx^\nu ,\notag\\
	&F^g_{\mu\nu}(x) = \kappa\{ \partial_\mu {\rm tr}(\bm{m}(x)\mathscr{A}_\nu(x)) - \partial_\nu {\rm tr}(\bm{m}(x)\mathscr{A}_\mu(x)) \notag\\
	&\qquad \qquad+ ig_{{}_{\rm YM}}^{-1}\sum_{i_1,\dots,i_n} \eta_{i_1\cdots i_n} {\rm tr}(\bm{m}(x)[\partial_\mu\bm{n}_{i_1}(x),[\bm{n}_{i_2},\cdots,[\bm{n}_{i_{n-1}}(x),\partial_\nu\bm{n}_{i_n}(x)]\cdots]])\}.
	\label{NAST-gen}
\end{align}
The detail of the derivation of (\ref{NAST-gen}) is given in Appendix \ref{sec:dnast}, which is almost the same as that given in \cite{MK15} for $SU(N)$, once 
(\ref{42}) is established.

\subsection{$G(2)$ case}

In each case of representations, we can write the new form for the Wilson loop operator using the decomposition formula based on the above general consideration:
\begin{enumerate}
\item
For $\tilde{H} = U(2)\in SU(3)$, the Wilson loop operator is written as (\ref{NAST-gen}) where $n=6$ and
\begin{align}
	\eta_{j_1\cdots j_6} = \zeta_{j_1\cdots j_6}
\end{align}
where $\zeta_{j_1\cdots j_6}$ is defined in (\ref{zeta}).

\item
For $\tilde{H} = U(2)\not\in SU(3)$, the Wilson loop operator is written as (\ref{NAST-gen}) where $n=6$ and
\begin{align}
	\eta_{j_1\cdots j_6} = \zeta'_{j_1\cdots j_6}
\end{align}
where $\zeta'_{j_1\cdots j_6}$ is defined in (\ref{zeta'}).


\item
For $\tilde{H} =U(1)\times U(1)$, the Wilson loop operator is written as (\ref{NAST-gen}) where $n=6$ and
\begin{align}
	\eta_{j_1\cdots j_6} = \zeta''_{j_1\cdots j_6}
\end{align}
where $\zeta''_{j_1\cdots j_6}$ is defined in (\ref{zeta''}).

By using $\bm{m} = \Lambda_i\bm{n}_i = (2n+m)\bm{n}_1/2 + m\bm{n}_2/(2\sqrt3)$, another form is obtained as
\begin{align}
	&F^g_{\mu\nu}(x) = \frac{2n+m}2 F^{(1)}_{\mu\nu}(x) + \frac m{2\sqrt3} F^{(2)}_{\mu\nu}(x) 
\notag\\
	&F^{(1)}_{\mu\nu}(x) = \kappa\{ \partial_\mu {\rm tr}(\bm{n}_1(x)\mathscr{A}_\nu(x)) - \partial_\nu {\rm tr}(\bm{n}_1(x)\mathscr{A}_\mu(x)) \notag\\
	&\qquad \qquad+ ig_{{}_{\rm YM}}^{-1}\sum_{j_1,\dots,j_6} \zeta'_{j_1\cdots j_6} {\rm tr}(\bm{n}_1(x)[\partial_\mu\bm{n}_{j_1}(x),[\bm{n}_{j_2},\cdots,[\bm{n}_{j_{5}}(x),\partial_\nu\bm{n}_{j_6}(x)]\cdots]])\} \notag\\
	&F^{(2)}_{\mu\nu}(x) = \kappa\{ \partial_\mu {\rm tr}(\bm{n}_2(x)\mathscr{A}_\nu(x)) - \partial_\nu {\rm tr}(\bm{n}_2(x)\mathscr{A}_\mu(x)) \notag\\
	&\qquad \qquad+ ig_{{}_{\rm YM}}^{-1}\sum_{j_1,\dots,j_6} \zeta_{j_1\cdots j_6} {\rm tr}(\bm{n}_2(x)[\partial_\mu\bm{n}_{j_1}(x),[\bm{n}_{j_2},\cdots,[\bm{n}_{j_{5}}(x),\partial_\nu\bm{n}_{j_6}(x)]\cdots]])\}.
\end{align}

\end{enumerate}

\section{Magnetic monopoles}

We can define magnetic-monopole current $k$ as the co-differential of the Hodge dual of $F^g$:
\begin{align}
	k = \delta^* F^g.
\end{align}
In the $D$-dimensional spacetime, $k$ is expressed by a differential form, $(D-3)$-form. 
For $D=4$, especially, the magnetic monopole current reads
\begin{align}
	k^\mu = \frac12\epsilon^{\mu\nu\rho\sigma}\partial_\nu F^g_{\rho\sigma}.
\end{align}
Then, the magnetic charge $q_m$ is defined by
\begin{align}
	q_m := \int d^3x k^0 = \int d^3x \frac12\epsilon^{jkl}\partial_l F^g_{jk} = \int d^2S_l \epsilon^{jkl}F^g_{jk}.
\end{align}
We examine the quantization condition for the magnetic charge.
The magnetic charge can have nonzero value because the map defined by
\begin{align}
	\bm{m}: S^2 \rightarrow G(2)/\tilde{H} = \begin{cases}
		G(2)/U(2) \\
		G(2)/(U(1)\times U(1))
	\end{cases},
\end{align}
has the nontrivial homotopy group:
\begin{align}
\label{77}
	\pi_2(G(2)/\tilde{H}) = \pi_1(\tilde{H}) =\begin{cases}
		\pi_1(S(2)\times U(1)) = \pi_1(U(1)) = \mathbb{Z}\\
		\pi_1(U(1)\times U(1)) = \mathbb{Z}+ \mathbb{Z}
	\end{cases}.
\end{align}
Because the value of the magnetic charge  depends only on the topological character of $\bm{n}_i$, we can use specific group elements $g$ to obtain the quantization condition for the magnetic charge. 
Now, we consider a case in which  $g(x)$ belongs to $SU(3)$.
In this case, $F^g_{\mu\nu}$ reduces to 
\begin{align}
	 F^g_{\mu\nu} =& \frac{2n+m}2F^{(1)}_{\mu\nu} + \frac m{2\sqrt3}F^{(2)}_{\mu\nu},\notag\\
	 F^{(1)}_{\mu\nu} =& \kappa\{\partial_\mu{\rm tr}(\bm{n}_1\mathscr{A}_\nu) - 2\partial_\nu{\rm tr}(\bm{n}_1\mathscr{A}_\mu)
	- ig_{{}_{\rm YM}}^{-1} {\rm tr}(\bm{n}_1[\partial_\mu\bm{n}_1,\partial_\nu\bm{n}_1] +\bm{n}_1[\partial_\mu\bm{n}_2,\partial_\nu\bm{n}_2])\}, \notag\\
	 F^{(2)}_{\mu\nu} =& \kappa\{\partial_\mu{\rm tr}(\bm{n}_2\mathscr{A}_\nu) - 2\partial_\nu{\rm tr}(\bm{n}_2\mathscr{A}_\mu)
	- ig_{{}_{\rm YM}}^{-1} {\rm tr}(\bm{n}_2[\partial_\mu\bm{n}_1,\partial_\nu\bm{n}_1] +\bm{n}_2[\partial_\mu\bm{n}_2,\partial_\nu\bm{n}_2])\} 
\nonumber\\  
=& \kappa \left\{\partial_\mu{\rm tr}(\bm{n}_2\mathscr{A}_\nu) - 2\partial_\nu{\rm tr}(\bm{n}_2\mathscr{A}_\mu)
	- \frac{4}{3} ig_{{}_{\rm YM}}^{-1} {\rm tr}( \bm{n}_2[\partial_\mu\bm{n}_2,\partial_\nu\bm{n}_2]) \right\}.
\end{align}
Here notice that two field strengths $F^{(1)}_{\mu\nu}$ and $F^{(2)}_{\mu\nu}$ appear in the non-Abelian Stokes theorem for $SU(3)$.
It is shown \cite{Kondo08,KKSS15} that the two kinds of the gauge-invariant charges $q_m^{(1)}$ and $q_m^{(2)}$ obey the different quantization conditions:
\begin{align}
	&q_m = \frac{2n+m}2 q_m^{(1)} +\frac m{2\sqrt3} q_m^{(2)}, \notag\\
	&q_m^{(1)} := \int d^3x \frac12\epsilon^{jk\ell}\partial_\ell F^g_{jk}  = \frac{4\pi}{g_{{}_{\rm YM}}}\left( \ell-\frac12 \ell'\right) ,\notag\\
	&q_m^{(2)} := \int d^3x \frac12\epsilon^{jk\ell}\partial_\ell F^g_{jk}  = \frac{4\pi}{g_{{}_{\rm YM}}}\frac12 \sqrt3\ell',\quad \ell,\ell'\in\mathbb{Z}. 
\end{align}
Thus, we obtain the quantization condition for the magnetic charge in $G(2)$:
\begin{align}
	q_m = \frac{4\pi}{g_{{}_{\rm YM}}}\left(\frac n2 (2\ell-\ell') + \frac m2 \ell\right) = \frac{2\pi}{g_{{}_{\rm YM}}} (n k +  m\ell),
\end{align}
where we have defined $k:= 2\ell-\ell'$, which can take an arbitrary integer.
The observation based on the homotopy group (\ref{77}) that there need to be two integers in $q_m$.
There exist already two integers in $q_m$.
Therefore, it is enough to consider a case $g(x) \in SU(3)$ for deriving the quantization condition for the magnetic charge in $G(2)$.

\section{Conclusions and discussions}

For the exceptional group $G(2)$, we have first shown that there exist three cases of the maximal stability subgroup.
Then, we have derived the gauge-covariant decomposition formula which is written using the multiple commutators with the color direction fields,   in accord with each stability group. 
Moreover, we have obtained the non-Abelian Stokes theorem for the Wilson loop operator that is written in terms of the relevant color direction fields.
These results indicate that there exist three options for the reformulation of the $G(2)$ Yang-Mills theory.
In any option, we need the two kinds of color fields, since the two Cartan generators are inevitably required in the decomposition formula, in marked contrast to the minimal option of $SU(N)$ group. 
Nevertheless, each option would be utilized for  describing confinement of quarks in the relevant  representation of $G(2)$.
This is because the the non-Abelian Stokes theorem for the Wilson loop operator is attributed to the  decomposition formula available to a given representation.
This would be confirmed more explicitly when the reformulation is ready to be checked.

The method we have used in this paper for obtaining the decomposition formula would be so general that  the decomposition formula is written  for any semi-simple Lie group using the multiple commutators with the Cartan generators.
In addition, once the decomposition formula given in the above is obtained, we can  immediately obtain the expression of the Wilson loop operator written in terms of the color direction fields, because we derived the non-Abelian Stokes theorem in a general way.
This observation suggests that the reformulation of the Yang-Mills theory with an arbitrary semi-simple gauge group would be possible.

\begin{acknowledgements}

This work is  supported by Grants-in-Aid for Scientific Research (C) No.24540252 and (C) No.15K05042 from the Japan Society for the Promotion of Science (JSPS).

\end{acknowledgements}

\appendix
\section{Necessity of sextuple commutators in the decomposition formula for  $G(2)$}
\label{sec:a}

In section \ref{sec:decom}, we have seen that  the sextuple commutator is used to obtain the decomposition formula for  $G(2)$.
In this appendix we show that such a formula for  $G(2)$ cannot be obtained by taking the commutator less than six times.
Taking the commutator odd number of times, we obtain 
\begin{align}
	[H_{j_1},\cdots,[H_{j_{2n+1}},\mathscr{F}]\cdots] = \sum_{\alpha\in\mathcal{R}_+} (\alpha_{j_1}\cdots\alpha_{j_{2n+1}} \mathscr{F}_\alpha^* E_\alpha - \alpha_{j_1}\cdots\alpha_{j_{2n+1}} \mathscr{F}_\alpha E_{-\alpha} ).
\end{align}
In this expression the sign of the term $E_\alpha$ is different from the sign of the term $E_{-\alpha}$ and therefore this is not appropriate.
Thus we see that we just need to consider the cases of double and quadruple commutators.

First, we consider the case of double commutators.
We can decompose an arbitrary element $\mathscr{F}$ of the Lie algebra if and only if there are real numbers $k_1$, $k_2$ and $k_3$ that satisfy
\begin{align}
\label{a1}
	 k_1[H_1,[H_1,\mathscr{F}]] + k_2[H_1,[H_2,\mathscr{F}]] + k_3[H_2,[H_2,\mathscr{F}]] 
	 = \sum_{\alpha\in\mathcal{R}_+:E_{\pm\alpha}\notin\tilde{h}}(\mathscr{F}_\alpha^* E_\alpha + \mathscr{F}_\alpha E_{-\alpha}).
\end{align}
Using the Cartan decomposition of $\mathscr{F}$, we find that the left hand side  of (\ref{a1}) is equal to
\begin{align}
	\sum_{\alpha\in\mathcal{R}_+}((\alpha_1)^2k_1 + \alpha_1\alpha_2k_2 + (\alpha_2)^2k_3)(\mathscr{F}_\alpha^* E_\alpha + \mathscr{F}_\alpha E_{-\alpha}),
\end{align}
where we put $\alpha = (\alpha_1, \alpha_2)$.
Hence (\ref{a1}) is equivalent to
\begin{align}
	&(\alpha_1)^2k_1 + \alpha_1\alpha_2k_2 + (\alpha_2)^2k_3 = 1 \quad {\rm for}\;E_\alpha\notin\tilde{h}, \notag\\
	&(\alpha_1)^2k_1 + \alpha_1\alpha_2k_2 + (\alpha_2)^2k_3 = 0 \quad {\rm for}\;E_\alpha\in\tilde{h}. 
	\label{a4}
\end{align}
In the case of $\tilde{H}=U(2)\in SU(3)$, the three equations (\ref{a4}) for $\alpha^{(1)}$, $\alpha^{(2)}$ and $\alpha^{(3)}$ can be written in a matrix form as
\begin{align}
	\begin{pmatrix}
		\frac14 &-\frac{\sqrt3}4 &\frac34 \\
		1&0&0\\
		\frac14 &\frac{\sqrt3}4 &\frac34
	\end{pmatrix}
	\begin{pmatrix}k_1\\k_2\\k_3\end{pmatrix} = \begin{pmatrix}0\\1\\1\end{pmatrix}.
\end{align}
The solution of this equation is $k_1=k_2=0$, $k_3=4/3$.
This solution is consistent with the decomposition formula for $SU(3)$.
But, these values of $k_1$, $k_2$ and $k_3$ do not satisfy the equation (\ref{a4}) for $\alpha^{(4)}$, $\alpha^{(5)}$ and $\alpha^{(6)}$.
For example, the equation (\ref{a4}) for $\alpha^{(5)}$ is given by
\begin{align}
	\frac13 k_3 = 1 ,
\end{align}
which is however not satisfied by $k_3=4/3$.
Therefore, there is no solution for all of (\ref{a4}).

In the case of $\tilde{H} = U(2) \notin SU(3)$ and of $\tilde{H} = U(1)\times U(1)$, the equation (\ref{a4}) for $\alpha^{(1)}$, $\alpha^{(2)}$ and $\alpha^{(3)}$ reads
\begin{align}
	\begin{pmatrix}
		\frac14 &-\frac{\sqrt3}4 &\frac34 \\
		1&0&0\\
		\frac14 &\frac{\sqrt3}4 &\frac34
	\end{pmatrix}
	\begin{pmatrix}k_1\\k_2\\k_3\end{pmatrix} = \begin{pmatrix}1\\1\\1\end{pmatrix}.
\end{align}
The solution of this equation is $k_1=k_3=1$, $k_2=0$.
This solution is also consistent with the decomposition formula for $SU(3)$.
But this does not satisfy the equation (\ref{a4}) for $\alpha^{(4)}$, $\alpha^{(5)}$ and $\alpha^{(6)}$.
Therefore, also in this case, there are no solutions for all of (\ref{a4}).
Thus we confirm that there are no decomposition formulae using double commutators for all representations.

Next, we consider the case of quadruple commutators.
There exits the decomposition formulae if and only if there are real numbers $k_1$, $k_2$, $k_3$, $k_4$ and $k_5$ that satisfy
\begin{align}
	&k_1[H_1,[H_1,[H_1,[H_1,\mathscr{F}]]]] + k_2[H_1,[H_1,[H_1,[H_2,\mathscr{F}]]]] + k_3[H_1,[H_1,[H_2,[H_2,\mathscr{F}]]]] + k_4[H_1,[H_2,[H_2,[H_2,\mathscr{F}]]]] \notag\\
	&+ k_5[H_2,[H_2,[H_2,[H_2,\mathscr{F}]]]] 
	 = \sum_{\alpha\in\mathcal{R}_+:E_{\pm\alpha}\notin\tilde{h}}(\mathscr{F}_\alpha^* E_\alpha + \mathscr{F}_\alpha E_{-\alpha}).
\end{align}
This is equivalent to
\begin{align}
	(\alpha_1)^4 k_1 + (\alpha_1)^3\alpha_2 k_2 + (\alpha_1)^2(\alpha_2)^2 k_3 + \alpha_1(\alpha_2)^2 k_4 + (\alpha_2)^4 k_5 = 1 \quad {\rm for} E_\alpha\notin\tilde{h}, \notag\\
	(\alpha_1)^4 k_1 + (\alpha_1)^3\alpha_2 k_2 + (\alpha_1)^2(\alpha_2)^2 k_3 + \alpha_1(\alpha_2)^2 k_4 + (\alpha_2)^4 k_5 = 0 \quad {\rm for} E_\alpha\in\tilde{h}. \notag\\
\end{align}
The equivalent matrix form is given as
\begin{align}
	\frac1{16}\begin{pmatrix}1 & -\sqrt{3} & 3 & -{{3}^{\frac{3}{2}}} & 9\cr 16 & 0 & 0 & 0 & 0\cr 1 & \sqrt{3} & 3 & {{3}^{\frac{3}{2}}} & 9\cr 1 & \frac{1}{\sqrt{3}} & \frac{1}{3} & \frac{1}{{{3}^{\frac{3}{2}}}} & \frac{1}{9}\cr 0 & 0 & 0 & 0 & \frac{16}{9}\cr 1 & -\frac{1}{\sqrt{3}} & \frac{1}{3} & -\frac{1}{{{3}^{\frac{3}{2}}}} & \frac{1}{9}\end{pmatrix}	
	\begin{pmatrix}k_1\\k_2\\k_3\\k_4\\k_5\end{pmatrix} = \begin{pmatrix}0\\1\\1\\1\\1\\1\end{pmatrix},\quad 
	\begin{pmatrix}1\\1\\1\\1\\0\\1\end{pmatrix}\quad {\rm or} \quad
	\begin{pmatrix}1\\1\\1\\1\\1\\1\end{pmatrix}.
\end{align}
Calculating the rank of the matrix, we see that these equations do not have the solutions.
Thus we have confirmed that quadruple commutators are not enough to obtain the desired decomposition formula.

\section{Derivation of the non-Abelian Stokes theorem for general gauge group using (\ref{42})}
\label{sec:dnast}

From the fact (\ref{42}) that  the $\tilde{H}$ part of $g^\dag\partial_\mu\bm{m} g$ is vanishing, we have  
\begin{align}
	g^\dag\partial_\mu\bm{m} g &= (g^\dag\partial_\mu\bm{m} g)_{G/H} \notag\\
	&=  \sum_{i_1,\dots,i_n} \eta_{i_1\cdots i_n}[H_{i_1},\cdots,[H_{i_n},g^\dag\partial_\mu\bm{m}g]\cdots].
\end{align}
Multiplying both sides of this equation by $g$ from the left and by $g^\dag$ from the right, we obtain 
\begin{align}
	\partial_\mu \bm{m}(x) &= \sum_{i_1,\dots,i_n} \eta_{i_1\cdots i_n}[\bm{n}_{i_1}(x),\cdots,[\bm{n}_{i_n}(x),\partial_\mu\bm{m}(x)]\cdots] \notag\\
	&= \sum_{i_1,\dots,i_n} \eta_{i_1\cdots i_n}[\bm{n}_{i_1}(x),\cdots,[\bm{m}(x),\partial_\mu\bm{n}_{i_n}(x)]\cdots] \notag\\
	&= \sum_{i_1,\dots,i_n} \eta_{i_1\cdots i_n}[\bm{m},[\bm{n}_{i_1}(x),\cdots,[\bm{n}_{i_{n-1}},\partial_\mu\bm{n}_{i_n}(x)]\cdots]] \notag\\
	&= ig_{{}_{\rm YM}} [\mathscr{B}_\mu(x), \bm{m}(x)],
	\label{49}
\end{align}
where we have used $[\bm{n}_i,\partial_\mu\bm{m}] = [\bm{m},\partial_\mu\bm{n}_i]$ following from $\partial_\mu[\bm{n}_i,\bm{m}] = 0$ in the second equality, the commuting property (\ref{18a}) in the third equality, and we have introduced
\begin{align}
	\mathscr{B}_\mu(x) := ig_{{}_{\rm YM}}^{-1}\sum_{i_1,\dots,i_n} \eta_{i_1\cdots i_n}[\bm{n}_{i_1}(x),\cdots,[\bm{n}_{i_{n-1}}(x),\partial_\mu\bm{n}_{i_n}(x)]\cdots]. 
\end{align}
in last equality.
On the other hand, we find
\begin{align}
	\partial_\mu\bm{m}(x) = ig_{{}_{\rm YM}}[\Omega_\mu,\bm{m}(x)].
	\label{51}
\end{align}
Combining (\ref{49}) and (\ref{51}), we obtain
\begin{align}
	[\Omega_\mu,\bm{m}(x)] = [\mathscr{B}_\mu(x),\bm{m}(x)] .\label{b5}
\end{align}
Using this relation, we can rewrite the third term in $F^g_{\mu\nu}(x)$ as
\begin{align}
	ig_{{}_{\rm YM}} {\rm tr}(\bm{m}(x) [\Omega_\mu(x),\Omega_\nu(x)]) &= ig_{{}_{\rm YM}} {\rm tr}([\bm{m}(x), \Omega_\mu(x)]\Omega_\nu(x)) \notag\\
	&= ig_{{}_{\rm YM}} {\rm tr}([\bm{m}(x), \mathscr{B}_\mu(x)]\Omega_\nu(x)) \notag\\
	&= ig_{{}_{\rm YM}} {\rm tr}([ \Omega_\nu(x),\bm{m}(x)]\mathscr{B}_\mu(x)) \notag\\
	&= {\rm tr}(\partial_\nu\bm{m}(x)\mathscr{B}_\mu(x)) \notag\\
	&= ig_{{}_{\rm YM}}^{-1}\sum_{i_1,\dots,i_n} \eta_{i_1\cdots i_n} {\rm tr}(\partial_\nu\bm{m}(x)[\bm{n}_{i_1}(x),\cdots,[\bm{n}_{i_{n-1}}(x),\partial_\mu\bm{n}_{i_n}(x)]\cdots]) \notag\\
	&= ig_{{}_{\rm YM}}^{-1}\sum_{i_1,\dots,i_n} \eta_{i_1\cdots i_n} {\rm tr}([\partial_\nu\bm{m}(x),\bm{n}_{i_1}(x)][\bm{n}_{i_2},\cdots,[\bm{n}_{i_{n-1}}(x),\partial_\mu\bm{n}_{i_n}(x)]\cdots]) \notag\\
	&= -ig_{{}_{\rm YM}}^{-1}\sum_{i_1,\dots,i_n} \eta_{i_1\cdots i_n} {\rm tr}([\bm{m}(x),\partial_\nu\bm{n}_{i_1}(x)][\bm{n}_{i_2},\cdots,[\bm{n}_{i_{n-1}}(x),\partial_\mu\bm{n}_{i_n}(x)]\cdots]) \notag\\
	&= -ig_{{}_{\rm YM}}^{-1}\sum_{i_1,\dots,i_n} \eta_{i_1\cdots i_n} {\rm tr}(\bm{m}(x)[\partial_\nu\bm{n}_{i_1}(x),[\bm{n}_{i_2},\cdots,[\bm{n}_{i_{n-1}}(x),\partial_\mu\bm{n}_{i_n}(x)]\cdots]])\notag\\ 
	&= ig_{{}_{\rm YM}}^{-1}\sum_{i_1,\dots,i_n} \eta_{i_1\cdots i_n} {\rm tr}(\bm{m}(x)[\partial_\mu\bm{n}_{i_1}(x),[\bm{n}_{i_2},\cdots,[\bm{n}_{i_{n-1}}(x),\partial_\nu\bm{n}_{i_n}(x)]\cdots]]),
\end{align}
where we have used the cyclicity of the trace in first, third, sixth and eighth equality, (\ref{51}) in fourth equality, $[\partial_\nu\bm{m},\bm{n}_{i_1}] = -[\bm{m},\partial_\nu\bm{n}_{i_1}]$ following from $\partial_\nu[\bm{m},\bm{n}_{i_1}]=0$ in seventh equality and the fact that the first expression is anti-symmetric in $\mu$ and $\nu$ in the last equality.
This completes the proof of the non-Abelian Stokes theorem (\ref{NAST-gen}). 

\section{Derivation of $F^g_{\mu\nu} = \kappa{\rm tr}(\bm m\mathscr F_{\mu\nu}[\mathscr V])$}
\label{sec:c}
In this appendix, we show that the remaining field $\mathscr X$ does not contribute to the Wilson loop operator by deriving the equality ($\ref{v}$).
Using the decomposition (\ref{decom_cb}), we obtain
\begin{align}
	{\rm tr}(\bm m\mathscr F_{\mu\nu}[\mathscr V])
	&= {\rm tr}(\bm m(\partial_\mu\mathscr C_\nu -\partial_\nu\mathscr C_\mu 
	-ig_{{}_{YM}}[\mathscr C_\mu, \mathscr C_\nu]
	-ig_{{}_{YM}}[\mathscr C_\mu, \mathscr B_\nu]
	-ig_{{}_{YM}}[\mathscr B_\mu, \mathscr C_\nu] \notag\\
	&\quad+ \partial_\mu\mathscr B_\nu -\partial_\nu\mathscr B_\mu 
	-ig_{{}_{YM}}[\mathscr B_\mu, \mathscr B_\nu] ).
\end{align}
From the fact (\ref{cm}) we see that the third, fourth and fifth terms vanish. Thus we obtain
\begin{align}
	{\rm tr}(\bm m \mathscr F_{\mu\nu}[\mathscr V]) = {\rm tr}(\bm m(\partial_\mu\mathscr C_\nu -\partial_\nu\mathscr C_\mu 
	+\partial_\mu\mathscr B_\nu - \partial_\nu\mathscr B_\mu -ig_{{}_{\rm YM}}[\mathscr B_\mu,\mathscr B_\nu])).\label{c1}
\end{align}
The first term of (\ref{c1}) reads
\begin{align}
	\partial_\mu{\rm tr}(\bm m\mathscr A_\nu) &= \partial_\mu({\rm tr}(\bm m\mathscr C_\nu)) \notag\\
	&= {\rm tr}(\partial_\mu\bm m\mathscr C_\nu + \bm m\partial_\mu\mathscr C_\nu) \notag\\
	&={\rm tr}(\bm m\partial_\mu\mathscr C_\nu), \label{c2}
\end{align}
where we have used $(g^\dag\partial_\mu\bm m g )_{\tilde H} =0$
and $\mathscr C_\nu \in g\;{\rm Lie}(\tilde H)g^\dag$.
The third term of (\ref{c1}) reads
\begin{align}
	{\rm tr}(\bm m\partial_\mu\mathscr B_\nu) &= -{\rm tr}(\partial_\mu\bm m\mathscr B_\nu) \notag\\
	&= -ig_{{}_{\rm YM}}{\rm tr}([\mathscr B_\mu,\bm m]\mathscr B_\nu) \notag\\
	&= ig_{{}_{\rm YM}}{\rm tr}(\bm m[\mathscr B_\mu,\mathscr B_\nu]), \label{c3}
\end{align}
where we have used ${\rm tr}(\bm m\mathscr B_\nu)=0$ in the first equality, (\ref{49}) in the second equality and the cyclicity of the trace in the last equality.
Thus we obtain
\begin{align}
	(\ref{c1}) &= \partial_\mu{\rm tr}(\bm m\mathscr A_\nu)- \partial_\nu{\rm tr}(\bm m\mathscr A_\mu)
	+ig_{{}_{\rm YM}} {\rm tr}(\bm m[\mathscr B_\mu, \mathscr B_\nu]) \notag\\
	&= \partial_\mu{\rm tr}(\bm m\mathscr A_\nu) - \partial_\nu{\rm tr}(\bm m\mathscr A_\mu)
	+ig_{{}_{\rm YM}} {\rm tr}(\bm m[\Omega_\mu, \Omega_\nu])\notag\\
	&= \frac1\kappa F^g_{\mu\nu},
	\end{align}
where we have used (\ref{c2}) and (\ref{c3}) in the first equality and (\ref{b5}) in the second equality.

\end{document}